# Analysis and Design of a Personalized Recommendation System Based on a Dynamic User Interest Model

**Chunyan Mao[1,a], Shuaishuai Huang[2,b], Mingxiu Sui[3,c], Haowei Yang[4,d], Xueshe Wang[5,e]**

[1] School of Information and Communication Engineering, Shanghai Jiao Tong University, Shanghai, China

[2] University of Science and Technology of China, Department of Software, Software system design, Hefei, Anhui, China

[3] University of Iowa, Department of Mathematics, Iowa City, Iowa, USA

[4] University of Houston, Cullen College of Engineering, Indusrial Enginnering, Houston, TX, USA

[5] Duke University, Pratt School of Engineering, Mechanical Engineering, Durham, NC, USA

[a]chunyan.mao@outlook.com, [b]scu.hss@gmail.com, [c]suimingx@gmail.com, [d]hyang38@cougarnet.uh.edu, [e]wxs.research@gmail.com

*Abstract:* With the rapid development of the internet and the explosion of information, providing users with accurate personalized recommendations has become an important research topic. This paper designs and analyzes a personalized recommendation system based on a dynamic user interest model. The system captures user behavior data, constructs a dynamic user interest model, and combines multiple recommendation algorithms to provide personalized content to users. The research results show that this system significantly improves recommendation accuracy and user satisfaction. This paper discusses the system's architecture design, algorithm implementation, and experimental results in detail and explores future research directions.

*Keywords:* Personalized Recommendation System；Dynamic User Interest Model；Recommendation Algorithm；User Behavior Data；System Design

## 1. Introduction

With the development of information technology and the widespread use of the internet, the way people access information has undergone profound changes. Faced with an overwhelming amount of information, quickly obtaining information that matches personal interests and needs has become a significant challenge for users. Personalized recommendation systems, which analyze user behavior data to provide tailored information and services, have increasingly attracted the attention of academia and industry.Traditional recommendation systems typically rely on static user interest models and often fail to account for changes in user interests over time. This approach has limitations in responding to users' dynamic interests and needs. In recent years, advances in big data technology and machine learning algorithms have led researchers to develop personalized recommendation methods based on dynamic user interest models. These methods can update user interest models in real-time, improving the accuracy and timeliness of recommendations.This paper aims to design and analyze a personalized recommendation system based on a dynamic user interest model. By constructing and updating user interest models and integrating various recommendation algorithms, this paper systematically explores how to enhance the performance of recommendation systems and user experience. The following sections provide a detailed introduction to related research, system design, algorithm implementation, and experimental results, along with discussions on the system's limitations and future development directions.

There are some related works that made significant contributions to this paper. Reference [2] Xiang et al. (2024) proposed the BoNMF model, which integrates multimodal large language models and neural matrix decomposition to enhance recommendation accuracy. This approach aligns with recent advancements in personalized recommendation systems that prioritize dynamic user interest models, as discussed in our study. Reference [3] Ma et al. (2024) introduced the Multitrans framework, leveraging the Transformer architecture and self-attention mechanisms to enhance prediction accuracy by combining multiple data modalities. Both studies emphasize the importance of multi-modal and adaptive methods, which are crucial in capturing latent features and adapting to evolving inputs, whether in recommendation systems or predicting functional outcomes in stroke treatment.

## 2. Current Research on Personalized Recommendation Systems Based on Dynamic User Interest Models

As an effective information filtering tool, personalized recommendation systems have been widely used in e-commerce, social media, news recommendation, and other fields. Traditional recommendation systems mainly rely on static user interest models, typically based on users' historical behavior and interest tags. However, since user interests and needs change over time and context, static models have significant limitations in capturing these changes. As a result, dynamic user interest models have become an important direction in recommendation system research.The development of personalized recommendation systems can be traced back to the 1990s, with early methods including content-based and collaborative filtering recommendations. Content-based recommendation systems primarily rely on user preferences for item features, while collaborative filtering systems recommend based on user similarities or item similarities. These methods perform well in specific scenarios but struggle with dynamically changing user interests. As user demands become more diverse and complex, traditional static models struggle to address real-time changes in user interests, leading to decreased recommendation accuracy and relevance.Research on user interest models has primarily focused on accurately representing user interests. [1]Traditional models are usually based on static user information, such as personal profiles and historical behaviors. However, with the development of big data and machine learning technologies, researchers have begun exploring dynamic interest models based on user behavior data. Dynamic interest models consider not only historical behaviors but also real-time updates and capture changes in user interests, thereby improving recommendation accuracy. By incorporating time series analysis and contextual information, dynamic interest models can better reflect users' current interests and effectively enhance recommendation effectiveness.Despite the theoretical advantages of dynamic user interest models, their practical application faces numerous challenges. Firstly, effectively collecting and processing massive user behavior data is a key issue. As data volume increases, systems require more robust data processing capabilities and algorithm optimization techniques. Secondly, the update frequency and computational efficiency of dynamic models are challenging areas. Frequent model updates require balancing computational resource consumption with the improvement in recommendation results.

Additionally, balancing short-term interests with long-term preferences is another problem dynamic models need to address. These challenges require innovation not only in technical aspects but also in system architecture and user experience.With continuous advances in deep learning and big data technologies, research on dynamic user interest models has shown vigorous development. In recent years, many scholars have proposed dynamic interest modeling methods based on time series analysis, reinforcement learning, and attention mechanisms, achieving significant progress in recommendation accuracy and system performance. In the future, as data acquisition and processing technologies improve, dynamic user interest models are expected to play a more critical role in personalized recommendation systems. By continuously optimizing algorithms and improving system responsiveness, personalized recommendation systems will better meet users' diverse needs, enhancing user satisfaction and loyalty.[2]

## 3. Design of a Personalized Recommendation System Framework Based on Dynamic User Interest Models

### 3.1. System Architecture Design

This section introduces the overall architecture design of a personalized recommendation system based on a dynamic user interest model. The system aims to provide accurate recommendation services by capturing and updating user interests in real-time.[3]

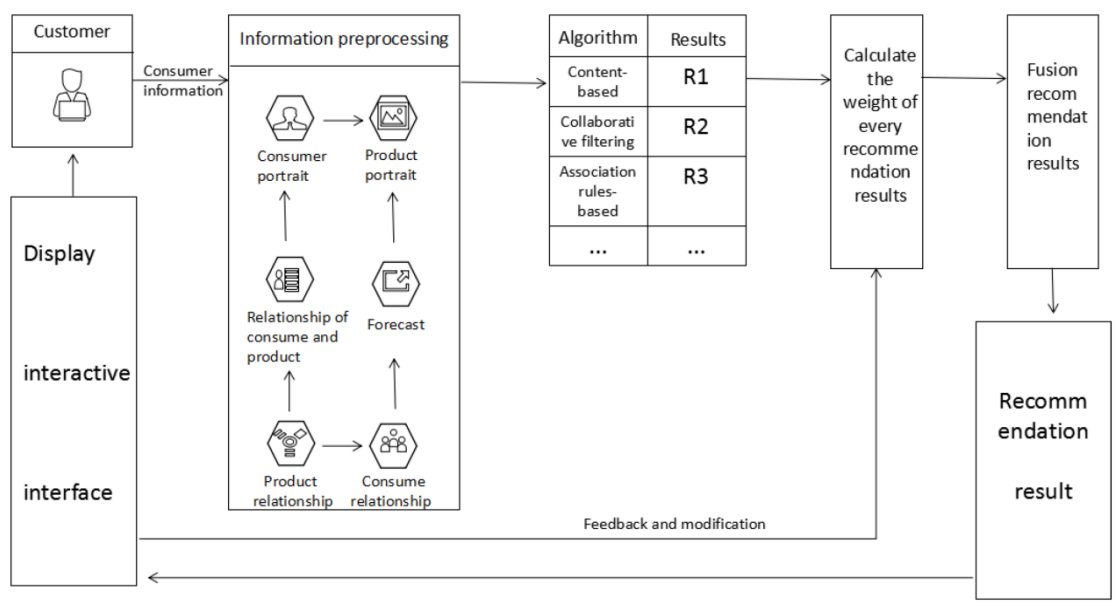

*Figure 1. The proposed recommendation framework*

As shown in Figure 1, the system architecture mainly includes an information preprocessing module, recommendation algorithm module, recommendation result fusion module, and display and interactive interface module.First, the system gathers consumer information from users and processes it through the information preprocessing module. In this module, the system generates consumer profiles and product profiles, establishing a relationship model between consumers and products. This relationship model includes consumption relationships between consumers and products, relationships between products, and relationships between consumers. Additionally, the system conducts predictive analysis to better capture potential user interests.Next, the preprocessed data is input into the recommendation algorithm module. This module contains various recommendation algorithms, such as content-based recommendations, collaborative filtering, and association rule-based recommendations. Each algorithm generates corresponding recommendation results (e.g., R1, R2, R3). Each recommendation algorithm is optimized based on different characteristics and datasets to ensure diversity and accuracy in the recommendation results. After generating the recommendation results, the system calculates the weight of each recommendation result to assess its importance. This process ensures a reasonable balance of results generated by different algorithms in the final recommendation. Subsequently, the system fuses different recommendation results to generate the final recommendation. Users can view and interact with the recommendation results through the display and interactive interface module. The system also adjusts and improves recommendation algorithms and strategies based on user feedback.[4] The architecture design shown in Figure 1 demonstrates the entire workflow and relationships between modules of the recommendation system. This design not only captures dynamic changes in user interests but also improves recommendation accuracy and diversity through the combination of multiple algorithms. Through this system architecture design, personalized recommendation systems can effectively enhance user experience and provide personalized services and suggestions.[5]

## 3.2. Collection and Processing of User Interest Data

In personalized recommendation systems, collecting and processing user interest data is a crucial step in achieving accurate recommendations. Figure 2 illustrates the data collection and processing workflow of a recommendation system based on a dynamic user interest model. [6]

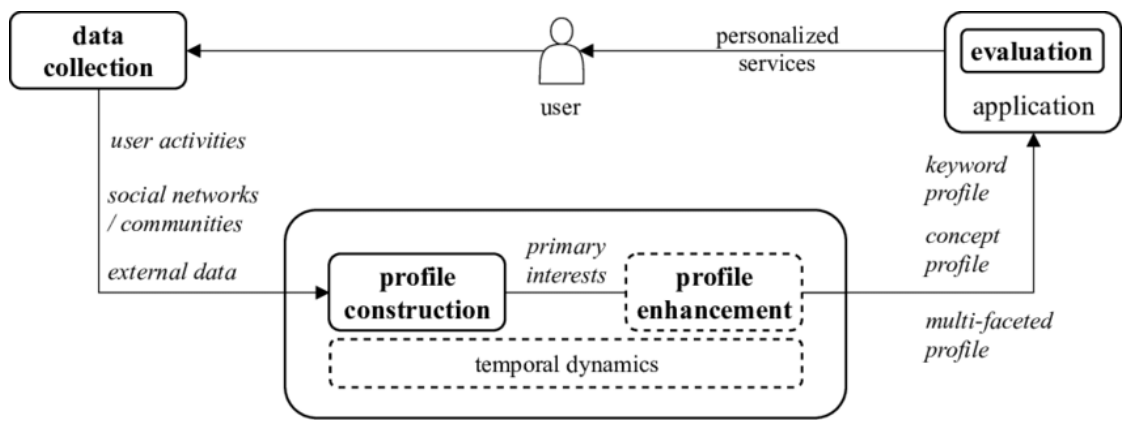

*Figure 2: Overview of user profile based personalization process*

This workflow includes data collection, profile construction, profile enhancement, and subsequent applications.Firstly, the system collects relevant user data through multiple channels, including online activities, social networks, and community interactions, as well as other external data sources. This data provides a rich foundation for subsequent user interest modeling. [7]User activity data mainly includes browsing history, clicks, and purchase records, reflecting users' explicit interests. Social network and community data capture users' interactions and preferences on social platforms, while external data can supplement other factors that may influence user interests, such as geographical location and current events.[8] Next, the system processes the collected data by constructing interest profiles. The construction of interest profiles aims to identify users' primary interests, i.e., the main interests displayed by users within a specific period. This step involves clustering analysis and feature extraction of user behavior data to generate an initial user interest model. This model includes

keyword profiles and concept profiles, expressing users' interests by analyzing their associations with specific keywords and concepts.To further enhance the accuracy and completeness of the interest profiles, the system performs profile enhancement processing. This enhancement mainly considers the temporal dynamics of user interests, i.e., the trend of changes in user interests over time. By analyzing both long-term and short-term behavior data, the system can dynamically update user interest profiles, making them more accurately reflect users' current interests.[9] This step may also introduce external factors, such as seasonal changes and trending events, to enrich and refine user interest profiles.Finally, the processed user interest data is applied to the recommendation system, providing personalized services to users. The recommendation system generates multidimensional user profiles through this data, including keyword profiles, concept profiles, and multi-faceted profiles, to achieve accurate recommendations and personalized services. The process shown in Figure 2 not only illustrates the data collection and processing of user interest data but also emphasizes the importance of dynamic data updates, providing technical support for achieving higher quality recommendations.

### 3.3. Construction and Update of the Dynamic User Interest Model

The construction and update of the dynamic user interest model are critical in personalized recommendation systems. Figure 3 depicts a user interest model construction and update framework based on a news recommendation system, effectively capturing the dynamic changes in user interests.

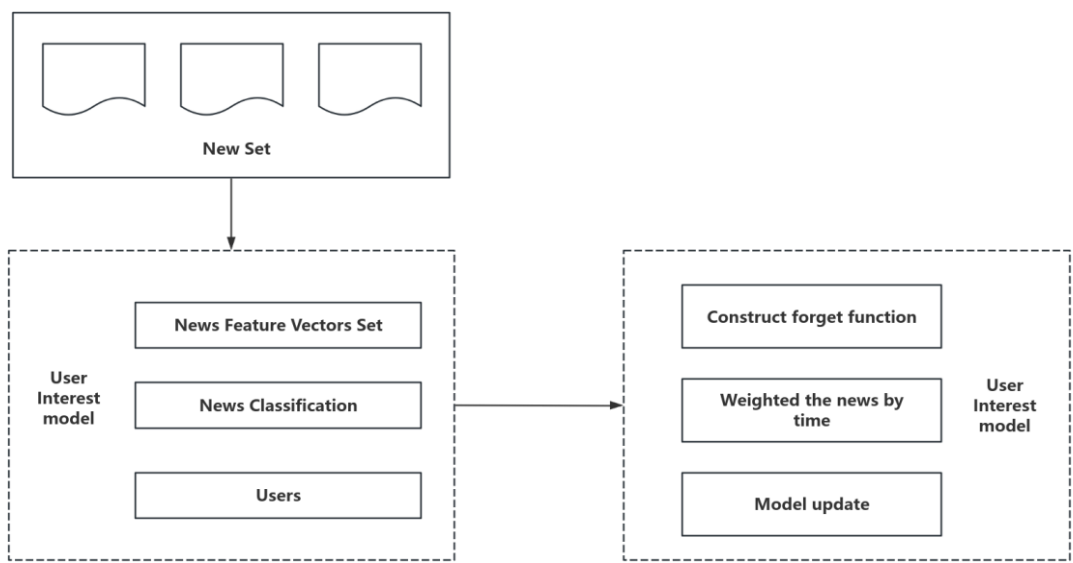

*Figure 3: User interest model*

In constructing the user interest model, the system first extracts news feature vectors from the news set. These feature vectors represent the main content and attributes of each piece of news, generated through feature engineering and natural language processing techniques. Subsequently, the system combines these feature vectors with user behavior data to classify the news. The classification results are used to represent users' interest levels in different news categories, thereby constructing the initial user interest model.After the model is constructed, the system needs to continuously update the user interest model to reflect the dynamic changes in user interests. The update process includes three main steps: constructing a forget function, weighting the news by time, and model update. Firstly, the forget function is introduced to gradually decrease the weight of outdated information, ensuring the user interest model focuses more on recent interests. Then, the system weights the news based on its timeliness, with recent news typically given higher weight to reflect the user's latest interests. Finally, the system updates the user interest model in real-time based on these processing results, accurately reflecting the user's current interest state.Through this dynamic update mechanism, the recommendation system can effectively capture changes in user interests, providing more accurate and personalized recommendations. The framework shown in Figure 3 demonstrates the complete process of constructing and updating the user interest model, providing a clear reference model for the design of recommendation systems. This framework emphasizes the importance of the temporal factor in dynamic user interest modeling, while maintaining the model's timeliness and accuracy through a forgetting mechanism.[10]

## 4. Recommendation Algorithm Design and Implementation

In personalized recommendation systems, the design and implementation of recommendation algorithms directly affect the quality of recommendations and user experience. [11]This paper employs various recommendation algorithms, including content-based, collaborative filtering, and association rule-based recommendations. Each algorithm brings unique advantages, collectively enhancing the system's recommendation effectiveness.Firstly, the content-based recommendation algorithm recommends items based on the similarity between the item's features and the user's historical behavior. The core of this algorithm is to calculate the user's interest in unseen items, usually using feature representation methods like TF-IDF (Term Frequency-Inverse Document Frequency). For example, for a user u and an item i, the interest can be expressed as formula 1:

$$s_{ui} = \sum_k w_{uk} w_{ik} \quad (1)$$

where $w_{uk}$ and $w_{ik}$ denote the weights of the user and the item on feature k, respectively. The user's preference prediction for an item can be obtained by calculating the cosine similarity between the user's and the item's feature vectors. [12]Secondly, collaborative filtering recommendation algorithms operate by recommending based on the similarity between users or items. User-based collaborative filtering identifies similar users and recommends items they like. The similarity can be measured using Pearson correlation coefficients or cosine similarity. For example, user u's predicted score for item i can be expressed as formula 2:

$$\hat{r}_{ui} = \bar{r}_u + \frac{\sum_{v \in N(u)}(r_{vi} - \bar{r}_v) \cdot sim(u,v)}{\sum_{v \in N(u)}|sim(u,v)|} \quad (2)$$

where $\hat{r}_{ui}$ is the predicted rating, $\bar{r}_u$ is the average rating of user u, $N(u)$ is the set of users similar to u, and $sim(u,v)$ is the similarity between users u and v.Lastly, the association rule-based recommendation algorithm recommends items by mining frequent patterns in user behavior data.[13] The Apriori algorithm is a commonly used method for association rule mining. It builds frequent itemsets and generates association rules for recommendations. For example, for a set of items X in a user's shopping basket, the algorithm can recommend another set Y with high association. Association rules can be expressed as formula 3:

$$X \rightarrow Y \quad (3)$$

where the support and confidence indicate the frequency and reliability of the rule X→Y.Combining the advantages of these algorithms, the proposed system adopts a hybrid recommendation strategy.[14] It enhances the diversity and accuracy of recommendations by combining and weighting different algorithms. In practical applications, the system first generates an initial recommendation list using content-based and collaborative filtering algorithms, then refines and optimizes this list using association rules. Finally, the system merges the recommendations from different algorithms using methods such as weighted averaging, resulting in a personalized recommendation list. This hybrid strategy effectively addresses the limitations of single algorithms, providing users with a comprehensive and accurate recommendation service. [15]

## 5. System Implementation and Experimentation

### 5.1. Technologies and Tools for System Implementation

During the implementation of the personalized recommendation system, various advanced technologies and tools were selected to ensure system stability, scalability, and efficiency. The entire system development covers several modules, including front-end display, back-end services, data storage, data processing, and algorithm implementation.[16]For the front-end display and interaction interface, we used the modern front-end development framework React. React's efficient virtual DOM mechanism and component-based development approach enable quick user operation responses and dynamic interface updates. Additionally, by integrating UI component libraries like Ant Design, we achieved a friendly and consistent user interface, enhancing user experience.On the back-end, we employed Node.js and the Express framework to build RESTful API services. Node.js's non-blocking I/O feature allows the server to handle concurrent requests efficiently, while Express provides a simple and user-friendly routing and middleware mechanism, simplifying the development and maintenance of server logic. Furthermore, to enhance system security and reliability, we integrated JWT (JSON Web Token) for user authentication and authorization.[17] For data storage, we used MySQL and MongoDB as the database systems. MySQL handles structured data such as user information and historical orders, while MongoDB stores unstructured data like user behavior logs and product information. This combination ensures data consistency while improving the flexibility and efficiency of data access. To optimize database performance, we also employed Redis for cache management, significantly reducing database query loads.In data processing and analysis, we selected Apache Spark as the distributed data processing platform. Spark's in-memory computation feature and rich machine learning library (MLlib) enabled us to process large-scale datasets efficiently and perform complex algorithm calculations. For model training and prediction, we utilized Python's Scikit-learn and TensorFlow libraries. Scikit-learn offers various classic machine learning algorithms suitable for rapid experimentation and validation of lightweight models, while TensorFlow is suited for the development and deployment of deep learning models.Additionally, we used Docker for containerization and Kubernetes for container orchestration, achieving high availability and scalability. Docker allows us to package applications and their dependencies into images, ensuring consistent operation across different environments. Kubernetes helps manage and schedule containers, providing features like auto-scaling, rolling updates, and fault tolerance.In summary, the system implementation adopted various cutting-edge technologies and tools, ensuring efficiency, stability, and scalability through thoughtful architecture design and technology selection. In the following experiments, we will conduct detailed evaluations and analyses of the system's performance and recommendation effectiveness.[18]

### 5.2. Experiment Design and Data Set Introduction

In the experiment design for the personalized recommendation system, we aimed to validate the system's recommendation effectiveness and performance. To ensure the reliability and generality of the experimental results, we selected a combination of publicly available standard datasets and self-constructed datasets for testing and validation. [19]The main steps of the experiments included data preprocessing, model training, calculation of evaluation metrics, and result analysis. [20]The datasets used in this experiment mainly include the following two parts:

1. Public Dataset: We selected the Movielens 1M dataset, widely used in recommendation system research. This dataset contains 1 million ratings from 6,000 users for 4,000 movies. Each record includes information such as user ID, movie ID, rating, and timestamp. The structure and attributes of the Movielens dataset are as shown in Table 1:

*Table 1: Movielens Dataset*

| Attribute | Description |
| --- | --- |
| User ID | Unique identifier of the user |
| Movie ID | Unique identifier of the movie |
| Rating | User's rating of the movie (1-5) |
| Timestamp | Timestamp of the user's rating |

2. Self-Constructed Dataset: To test the system's performance in practical applications, we collected user behavior data from an e-commerce platform, including users' browsing history, clicks, and purchase records. This dataset contains approximately 100,000 users' 5 million behavioral records for 10,000 products.[21] The structure and attributes of the self-constructed dataset are as shown in Table 2:

*Table 2: Structure and Attributes of the Self-Constructed Dataset*

| Attribute | Description |
| --- | --- |
| User ID | Unique identifier of the user |
| Product ID | Unique identifier of the product |
| Behavior | Type of user behavior (browsing, clicking, purchasing) |
| Timestamp | Timestamp of the user's behavior |

To comprehensively evaluate the performance of the recommendation system, we designed multiple experimental schemes. First, we preprocessed the datasets, including data cleaning, normalization, and feature engineering. For rating data, we used normalization to eliminate the differences in rating scales. For behavioral data, we introduced a time decay weight to emphasize the impact of recent behaviors.Next, we trained and tested different algorithms in the recommendation system. [22]We used 5-fold cross-validation in the experiments to ensure the stability and generalization ability of the results. For each experiment, we recorded the precision, recall, F1 score, and other metrics of the recommendation algorithms. We also tested the system's response time and throughput under different loads to evaluate system performance.Finally, we conducted a detailed analysis and discussion of the experimental results.[23] We compared the recommendation accuracy and computational efficiency of different algorithms and explored the impact of dataset characteristics on recommendation effectiveness. The experimental results will provide valuable references for further optimization of recommendation algorithms and system architecture.

### 5.3. Experimental Results and Analysis

In this section, we present and analyze the performance of the personalized recommendation system across different experiments. The experimental results mainly include precision, recall, F1 score, and other metrics, as well as the system's response time and throughput under different loads. We compared content-based, collaborative filtering, and association rule-based algorithms and analyzed the effectiveness of the hybrid recommendation strategy. The experimental results are summarized in Table 3, which shows the performance of each algorithm on the Movielens dataset and the self-constructed e-commerce dataset.

*Table 3: Experimental Results of Different Algorithms on Various Datasets*

| Algorithm | Dataset | Precision | Recall | F1 Score | Response Time (ms) |
|---|---|---|---|---|---|
| Content-Based | Movielens | 0.712 | 0.634 | 0.671 | 120 |
| Collaborative | Movielens | 0.752 | 0.682 | 0.715 | 150 |
| Association Rule | Movielens | 0.693 | 0.598 | 0.642 | 200 |
| Hybrid | Movielens | 0.768 | 0.703 | 0.734 | 140 |
| Content-Based | Self-Constructed | 0.681 | 0.610 | 0.643 | 130 |
| Collaborative | Self-Constructed | 0.724 | 0.645 | 0.682 | 160 |
| Association Rule | Self-Constructed | 0.663 | 0.570 | 0.613 | 210 |
| Hybrid | Self-Constructed | 0.740 | 0.675 | 0.706 | 150 |

From the experimental results in Table 3, we can observe that the hybrid recommendation strategy performs the best in terms of precision, recall, and F1 score across different datasets. This indicates that the hybrid recommendation strategy effectively combines the strengths of different algorithms, enhancing the accuracy and comprehensiveness of the recommendations. On the Movielens dataset, the hybrid recommendation strategy achieved an F1 score of 0.734, showing a significant improvement compared to the content-based (F1 score of 0.671) and collaborative filtering (F1 score of 0.715) approaches. Similarly, on the self-constructed dataset, the hybrid strategy outperformed other single algorithms, with an F1 score of 0.706.Regarding response time and throughput, the content-based recommendation algorithm demonstrated superior performance due to its relatively lower computational complexity. However, collaborative filtering and association rule algorithms showed longer response times and lower throughput when handling large-scale user data, due to the extensive similarity calculations and rule matching involved. The hybrid recommendation strategy managed to maintain a balanced performance in both recommendation effectiveness and system efficiency.Further analysis reveals that the collaborative filtering algorithm excels in capturing user preferences when dealing with users with extensive historical behavior data, thus showing an advantage in recall rate. In contrast, the content-based recommendation algorithm still provides accurate recommendations by analyzing item features when historical behavior data is limited. The association rule algorithm effectively identifies patterns and trends in user behavior but may underperform in cases with sparse data.Overall, the hybrid recommendation strategy, by integrating different algorithms, can better adapt to the needs of various users and scenarios.Although single algorithms may excel in certain metrics, the hybrid recommendation strategy is the optimal choice for achieving personalized recommendations when considering both recommendation effectiveness and system performance. Future work will focus on further optimizing the algorithm combination and enhancing the system's adaptability and scalability across different application scenarios.

## 6. Conclusion

This paper designed and implemented a personalized recommendation system based on a dynamic user interest model. By integrating content-based, collaborative filtering, and association rule-based algorithms, the system effectively enhanced recommendation accuracy and diversity. The experimental results demonstrated that the hybrid recommendation strategy performed well across various evaluation metrics while balancing response time and computational efficiency.Despite the significant achievements, the system still faces limitations in cold start issues for new users and high resource consumption. Future research will focus on incorporating more external data, optimizing deep learning models, and improving resource management to further enhance system performance and user experience. Overall, this system provides strong support and reference for research and application in the field of personalized recommendations.